# Anomalous Roughening of Curvature-Driven Growth With a Variable Interface Window

Yong-Jun Chen[1,3], Yuko Nagamine[2], Tomohiko Yamaguchi[3] and Kenichi Yoshikawa[1,2*]

[1]Department of Physics, Graduate School of Science, Kyoto University, Kyoto 606-8502, Japan

[2]Spatio-Temporal Order Project, ICORP, JST (Japan Science and Technology Agency), Kyoto 606-8502, Japan

[3]Nanosystems Research Institute, National Institute of Advanced Industrial Science and Technology (AIST), 1-1-1 Higashi, Tsukuba, Ibaraki, 305-8565, Japan

∗Corresponding author: yoshikaw@scphys.kyoto-u.ac.jp

Abstract

We studied the curvature-driven roughening of a disk domain pattern with a variable interface window. The relaxation of interface is driven by negative "surface tension". When a domain boundary propagates radially at a constant rate, we found that evolution of interface roughness follows scaling dynamic behavior. The local growth exponents are substantially different from the global exponents. Curvature-driven roughening belongs to a new class of anomalous roughening dynamics. However, a different "surface tension" leads to different global exponents. This is different from that of interface evolution with a fixed-size window, which has universal exponent. The variable growth window leads to a new class of anomalous roughening dynamics.

Introduction

Interface growth occurs in a variety of situations, including crystal growth [1], electrochemical deposition [2-10], directed percolation [11], displacement of one fluid by another [12], pattern formation [13] and bacterial cloning [14-16]. It has been shown that the roughening of an interface obeys a universal class of scaling behavior [17, 18]. The width of an interface, $W$, which is referred to as its roughness, grows according to a scaling law with respect to time and the size of the interface. Many authors have sought to explain the scaling behavior of interface growth [17-28]. A celebrated model for interface growth is the Kardar-Parisi-Zhang (KPZ) equation [17], which can predict roughening behavior. A dynamic scaling hypothesis has been proposed by Family and Viscek (FV) for a fixed growth window [28]: $W(L) \sim t^\beta$ for $t < \tau$, $W(L) \sim L^\alpha$ for $t > \tau$ and $\tau \sim L^z$, where $L$ is the fixed lateral size of the growth window used to measure width $W$, $t$ is time, $\tau$ is the saturation time of roughness, $\beta$ is a growth exponent, $\alpha$ is a roughness exponent, and $z = \alpha/\beta$ is a dynamic exponent. However, experiments under various conditions have revealed a rich variety of anomalous roughening scaling [3-10, 28]. For example, in electrochemical deposition [6-7], a series of local exponents, which are substantially different from global exponents, have been demonstrated. A generic kinetic theory based on a scaling analysis has been proposed by Ramasco *et al.* to classify the dynamic scaling behavior [18]. Though various effects have been discussed by the authors [6-10], the reason for the appearance of anomalous scaling is still not clear. Previous observations have particularly focused on the roughening process with a fixed-size window [3-10, 17-28]. Recently, it has been shown that dynamic scaling behavior is important for tumor therapy [29-31]. The edge of a tumor propagates radially [29], and the universal class of scaling to which tumor growth belongs is not yet clear [32, 33]. Since the interface window is not fixed and the size of the window changes with time, which has not been elucidated in previous scaling analysis, the universal class of dynamic scaling remains to be established. In this article, we present the anomalous roughening dynamics of curvature-driven growth, where the growth window is not fixed and the interface propagates radially. We propose that curvature-driven growth produces a new class of anomalous scaling dynamics. In addition, our results show that dynamic scaling depends on "surface tension" of the interface. In contrast to the roughening dynamics of an interface with a fixed-size window, a variable window provides new scaling dynamics of interface growth.

Theoretical model

In a simple case where the substrate is near flat and a particle is deposited ballistically on the planar substrate, interface growth can be described using the Edwards-Wilkinson (EW) equation [28]

$$\frac{\partial h}{\partial t} = \upsilon + \nu \nabla^2 h + \eta \qquad (1)$$

where $h$ is the position of the interface, $\upsilon$ is the mean velocity, $\nu$ is "surface tension" and $\eta = \eta(t)$ is random noise. Equation (1) provides a local equation of motion for a moving front [34]

$$\hat{n} \cdot \frac{d\vec{r}}{dt} = \overline{\upsilon}(\vec{r}) + \gamma \kappa + \eta \qquad (2)$$

where $\hat{n}$ is the normal of the moving front, $\vec{r}$ is a generalized coordinate on the moving front, $\overline{\upsilon}(\vec{r})$ is the mean normal velocity induced by the external force, $\gamma$ is a constant and $\kappa$ is the local curvature. However, under actual experimental conditions, the moving front often exhibits morphologic instability. For example, in the case of electrochemical deposition [6, 7], fluid displacement in a disordered medium [34], phase percolation [12], and spontaneous imbibition [35, 36], a moving front expands normally and the local velocity is normal to the interface. These experimental observations do not agree with the prediction based on EW equation [28]. While Eq. (1) is not valid, Eq. (2) is still sound when we consider the interface at a macroscopic scale [34]. Equation (2) has been successfully applied to describe the motion of a solidification front [37] and a phase boundary [38]. In addition, Eq. (2) can be derived from the Ginzburg-Landau equation for a phase transition [26]. The local curvature can be expressed as $\kappa = -\hat{n} \cdot \frac{\partial^2 \vec{r}}{\partial s^2}$, where $s$ is the arc length on the moving boundary. Considering that the interface expands normally and the velocity of interface is normal to the moving front, Eq. (2) can be expressed as

$$\frac{\partial \vec{r}}{\partial t} = \overline{\upsilon}(\vec{r})\hat{n} - \gamma \hat{n} \cdot \frac{\partial^2 \vec{r}}{\partial s^2} \hat{n} + \hat{n}\eta \qquad (3)$$

By decomposing coordinate $\vec{r}$ in Eq. (3), we obtain

$$\frac{\partial x}{\partial t} = \overline{\upsilon}(\vec{r})n_x - \gamma(\frac{\partial^2 x}{\partial s^2}n_x + \frac{\partial^2 y}{\partial s^2}n_y)n_x + \eta n_x \qquad (4)$$

$$\frac{\partial y}{\partial t} = \overline{\upsilon}(\vec{r})n_y - \gamma(\frac{\partial^2 x}{\partial s^2}n_x + \frac{\partial^2 y}{\partial s^2}n_y)n_y + \eta n_y \qquad (5)$$

where coordinate $\vec{r}(s)$ has been decomposed to $\vec{r} = x\vec{i} + y\vec{j}$, the normal direction on the curve $\hat{n}$ has been decomposed to $\hat{n} = n_x \hat{i} + n_y \hat{j}$, and, $\vec{i}$ and $\vec{j}$ are unit vectors in two orthogonal directions of $x$ axis and $y$ axis, respectively. In Eqs. 4 and 5, white noise ($\eta(x,t)$) is included to replicate the actual evolution of the interface in a disordered condition. For a suitable noise intensity, we found in the simulation that the intensity of white noise does not affect the growth exponent and roughness exponent. This model (curvature-driven Laplacian growth) is compatible with the EW equation at a microscopic scale [28]. However, the lateral growth induced by nonlinearity is also a dominant factor in the growth of the interface [28]. Experiments on electrochemical deposition [6-10] have provided clear evidence for lateral growth. Consequently, according to the KPZ theory, we obtain the following equations by adding a KPZ nonlinear term to Eqs. 4 and 5:

$$\frac{\partial x}{\partial t} = \overline{\upsilon}(\vec{r})n_x - \gamma(\frac{\partial^2 x}{\partial s^2}n_x + \frac{\partial^2 y}{\partial s^2}n_y)n_x + \frac{\lambda}{2}(\frac{\partial x}{\partial s})^2 \text{sgn}(n_x) + \eta n_x \qquad (6)$$

$$\frac{\partial y}{\partial t} = \overline{\upsilon}(\vec{r})n_y - \gamma(\frac{\partial^2 x}{\partial s^2}n_x + \frac{\partial^2 y}{\partial s^2}n_y)n_y + \frac{\lambda}{2}(\frac{\partial y}{\partial s})^2 \text{sgn}(n_y) + \eta n_y \qquad (7)$$

where $\lambda$ is constant and $\text{sgn}(\omega)$ is a sign function ($\text{sgn}(\omega) = 1$ when $\omega > 0$, $\text{sgn}(\omega) = -1$ when $\omega < 0$, otherwise $\text{sgn}(\omega) = 0$). Equations 6 and 7 are the formulation of KPZ-like equations. KPZ theory predicts that the width of a moving interface described by the KPZ equation obeys scaling law in the form $W \sim t^\beta$, where $W$ is the width of the interface, $t$ is time and $\beta = 1/3$ is a growth exponent for one-dimension growth [28]. Thus, we expect that the above model will produce scaling law of the interface width.

In the above equations (Eqs. 6 and 7), the velocity of an interface includes the mean velocity and the contribution from nonlinear effects [28]. The interface will deform and lateral growth will become important for roughening of the interface [28]. The positive effective "surface tension" prefers to flatten the interface, and the interface will grow asymptotically to a flat interface [17]. Thus, positive "surface tension" can not drive the growth of an interface. If we consider the free energy of the system, the growth of a domain pattern tends to decrease the free energy of the system. This implies that the effective "surface tension" becomes negative, $-\gamma < 0$, which will expand the interface in the normal direction. The driving force for growth is induced by the deposition of solute [3-10] or an imbalance of tension [38]. The normal growth of the interface in two dimensions can be decomposed into motions in the directions of the $x$ axis and $y$ axis. Therefore, we can expect that the growth can be described by combining Eq. 6 and 7.

Numerical Results and Discussion

Here, we consider a two-dimensional disk domain that is initially circular. The initial disorder is magnified and coarsened to a domain pattern (Fig. 1). In the simulation, we considered merging of the interface as in Ref. [38]. The smaller domain created in the coalescence of domain boundaries is neglected. Thus, there is a switch in the evolution of width of the domain boundary after the coalescence of domain boundaries. Small concavities and small protrusions have large values of curvature. A large curvature corresponds to a disordered state (curvature as an order parameter). A change in morphology under an external potential is similar to the phase transition from disorder to order. In the long term, the curvature will approach zero (ordered state). Thus, concavities and protrusions on the boundary of the domain will be coarsened. To test the roughening dynamics of an interface, we calculate the roughness of the interface, which is described as its width. The width of the interface is the root-mean-square of interface fluctuation

$$W(L,t) = <r^2 - \bar{r}^2>^{1/2}$$

where $r$ and $\bar{r}$ are the distance to the center of the pattern and its mean value, and $<>$ represents an average over the space. A log-log plot of the width of an interface against time is shown in Fig. 2. At an early time, the evolution of width follows a scaling form. We can express width as a scaling form, $W \sim t^\beta$, where the growth exponent $\beta$ is about $0.5$ as indicated by the slope of the guideline in Fig. 1. Since the evolution is driven by curvature, at late stage, the small curvature leads to roughness saturation as shown in Fig. 2. The width of interface will then decrease because the interface is being smoothed out and the size of the interface continues to changes. The time of saturation depends on the mean velocity because expansion of the domain will reduce the mean curvature. Remarkable fluctuation is found in the scaling regime and saturation regime (Fig. 2). We attribute this to the annihilation and creation (because of noise) of concavity and protrusions on the moving interface. The total trend before saturation does not depend on the mean velocity, as shown in Fig. 2. The growth exponent produced by our simulation is different from a previous prediction $\beta=1/3$ by the KPZ equation in one dimension [17]. In the plot of the width of the interface, a glitch is created by the merging of domain boundaries in some cases. "Surface tension" $-\gamma$ affects the evolution of the domain. The evolution of width shows the different direction and total trend for different values of "surface tension" $-\gamma$. We obtain different growth rate of roughness with different values of "surface tension" as shown in Fig. 3. The previous prediction by KPZ theory indicates that "surface tension" does not affect the dynamic scaling of interface evolution and thus the scaling is universal in one dimension with a fixed-size window [28]. The variable interface window provides a new class of interface evolution. However, if we set parameters $\gamma$ and $\lambda$ to zero, the evolution of the domain boundary will be identical to random deposition. Interestingly, random deposition also produces scaling behavior (Fig. 4). The predicted growth rate agrees with the result from simulation of random deposition on a one-dimension substrate [28]. This suggests that random deposition is to some extent a special curvature-driven growth without surface tension. According to the hypothesis of Family and Vicsek [28], the width of an interface has a scaling form:

$$W = L^\alpha f(t/L^z)$$

where $f(t/L^z)$ is a scaling function. The saturated roughness should follow a scaling form versus the size of the interface window when time is greater than a critical time for saturation. However, the size of the interface window ($L=2\pi\bar{r}$) changes with time in our system. It is not convenient to obtain a roughness exponent $\alpha$ by calculating the width of the interface at a different window size. Instead, it is better to calculate the structure function $S(k,t)=|r(k,t)|^2$, where $k$ is wave number and $r(k,t)=\int r(s,t)e^{iks}ds$ is a Fourier transformation of the radius. For a self-affine interface, the structure function has a scaling form [18]

$$S(k,t) = k^{-(2\alpha+1)} f_s(kt^{1/z}) \quad \text{with} \quad f_s(u) \sim \begin{cases} u^{2(\alpha-\alpha_s)} & \text{if} \quad u \gg 1 \\ u^{2\alpha+1} & \text{if} \quad u \ll 1 \end{cases} \qquad (8)$$

where $f_s(u)$ is a spectral scaling function and $\alpha_s$ is a spectral roughness exponent. Figure 5(a) shows a log-log plot of the structure function against wave number. As shown in Fig. 5(a), the slope of the guideline depends on time. We get the slopes of guidelines -4.4 at time t=0.40, -3.0 at time t=2.0, and -2.8 at time t=3.8, which do not indicate the same roughness exponent. It is difficult to determine the universal roughness exponent. To determine a universal roughness exponent, we plot scaling function ($f_s = S(k,t)k^{2\alpha+1}$) against ($kt^{1/z}$) in Fig. 5(b) using $\beta = 0.5$ and tentative roughness exponents ($\alpha = 0.8$, $\alpha = 1.1$ and $\alpha = 1.3$, respectively, as shown in Fig. 5(b)). The data collapses to a universal scaling at different points when roughness exponent is in the range of 1.0 to 1.25. Thus we estimate that the global roughness exponent $\alpha$ is approximately 1.10. From Fig. 5(b) we obtain a spectral roughness exponent $\alpha_s = 1.03$ (using $2(\alpha - \alpha_s) = 0.15$). The value of "surface tension" affects the evolution of structure function and thus affects the roughness exponent as shown in Fig. 6. The slope of evolution of structure function against wave number (log-log plot in Fig. 6) can be expressed as $-(2\alpha+1)$. Larger value of "surface tension" will lead to smaller roughness exponent as shown in Fig. 6. This is rather different from interface evolution with a fixed-size window, which has a universal roughness exponent [28]. However, for random deposition ($\gamma = 0$ and $\lambda = 0$), roughness exponent is 0.5 (Fig. 4(b)), which is identical to the result in the simulation of one-dimensional deposition on a straight substrate [28]. The geometry of the substrate for growth plays an important role in the evolution of roughness of the interface.

To test the local exponent of roughening dynamics, we calculate the local roughness and local structure function. The closed domain is divided into 12 parts by center angle ($\theta$) from the center of the domain (Fig. 7(a)). The local width of the interface and the local structure function are $W_{loc}(\theta,t) = <r^2 - \bar{r}^2>_\theta^{1/2}$ and $S_{loc}(k,t) = |r(k,t)|_\theta^2$, respectively. Figure 7 shows a plot of the local roughness against time (Fig. 7(b)) and the local structure factor against wave number for interfaces of different sizes (center angle from the center of the domain) (Fig. 7(c)). The roughness evolution and structure factor clearly depend on the size of the interface. We obtain the local growth exponent $\beta_{loc} = 0.25$ as shown by dashed guideline in Fig. 7(b) when center angle is $\pi/6$. And we obtain the local roughness exponent $\alpha_{loc} = 0.5$ when the center angle is $\pi/6$ as shown by the slope of guideline $-(2\alpha_{loc}+1)$ in Fig. 7(c). Though global roughness exponent depends on time and is not universal, the local roughness does not depend on time as shown by Fig. 7(c). When the local exponent is not zero, we obtain the relation $\beta_{loc} = (\alpha - \alpha_{loc})/z$ [35, 36]. Using above relation, we can get global roughness exponent $\alpha = 1.0$, which is agree with our above estimation when "surface tension" $-\gamma = -3.0$ (Fig. 5(b)). According to the generic classification of kinetic roughening [27], curvature-driven growth is *a new class of anomalous roughening* because $\alpha_{loc} \neq \alpha_s$ and $\alpha \neq \alpha_{loc}$ ($\alpha_{loc} \neq 1.0$).

Conclusion

In summary, we have shown the anomalous scaling dynamics of curvature-driven growth with a non-fixed growth window. The relation between global exponents and local exponents indicates that curvature-driven growth is a new class of anomalous roughening dynamics. Our simulation demonstrates that a different "surface tension" leads to different exponents. Thus, it has become clear that a non-fixed growth window produces a new class of roughening dynamics. Finally, we remark upon the effect of shape of interface window on the scaling dynamics of tumor growth [29]. As shown in our simulation, a variable interface window makes striking difference for the dynamics of interface propagation. For the universal class of the tumor dynamics, new scaling analysis should be considered instead of following the traditional scaling analysis based on the fixed-size planar window. And also the new method should be used to calculate local exponents as that in our present article (Fig. 7(a)). To simply connect the tumor dynamics with the universal class of linear molecular beam epitaxy model could be misleading [29]. Further consideration on the tumor dynamics is needed.

Acknowledgement

Yong-Jun Chen thanks the Ichikawa Foundation for support. Y. -J. C. received support from the Global COE program of Kyoto University. This work was supported by a Grant-in-aid for Creative Scientific Research (Project No. 18GS0421). This work was also partly supported by the Japanese Ministry of Education, Culture, Sports, Science and Technology (MEXT) via Grant-in-Aid for Scientific Research on Innovative Areas "Emergence in Chemistry" (20111007).

Figure captions

Fig. 1 (color online) Profile of moving boundary at different points. Parameters for the simulation: $\gamma = 3.0$, $\lambda = 2.0$, $\bar{\upsilon}(\vec{r}) = 1.0$. In all the simulation of this article, we choose white noise $\eta$ which is in the range of [-0.01, 0.01].

Fig. 2 (color online) Log-log plot of the global width of the boundary of the disk domain pattern against time at various mean velocities. The normal mean velocities $\bar{\upsilon}(\vec{r})$ of the boundary are 0, 0.1, 0.3, 0.6, 1.0, 2.0, and 3.0, respectively. Time step is $\Delta t = 0.0002$. The dashed line is a guide for the eyes. The slope of the dashed line is 0.5. Parameters for the simulation: $\gamma = 3.0$, $\lambda = 2.0$.

Fig. 3 (color online) Log-log plot of growth of interface roughness with different "surface tension". The values of "surface tension" $-\gamma$ are indicated in figure. Parameters for the simulation: $\lambda = 2.0$, $\bar{\upsilon}(\vec{r}) = 1.0$.

Fig. 4 (color online) Log-log plot of width and structure function when $\gamma = 0$, $\lambda = 0$. Parameters for the simulation: $\bar{\upsilon}(\vec{r}) = 1.0$.

Fig. 5 (color online) Log-log plot of the global structure function $S(k,t)$. (a) Structure function at different point, t=0.40, t=2.0, t=3.8, respectively. The slope of the guideline is -4.4, -3.0, and -2.8, respectively. (b) Collapsed rescaled structure function at different points using different global roughness exponent 0.80, 1.10 and 1.30, respectively. The dashed lines are guides for eyes. The dashed guidelines have the same slope of 0.15 when using $\alpha = 1.10$. Parameters for the simulation: $\gamma = 3.0$, $\lambda = 2.0$, $\bar{\upsilon}(\vec{r}) = 1.0$.

Fig. 6 (color online) Dependence of structure function on "surface tension". The dashed line is guidance for eyes. The slopes of guidelines are indicated in the figures. The values of "surface tension" are $\gamma = 1.0$, $\gamma = 2.0$, and $\gamma = 4.0$, respectively. Parameters for the simulation: $\lambda = 2.0$, $\bar{\upsilon}(\vec{r}) = 1.0$. To avoid overcrowding 1.0 is added to $\log_{10} S(k,t)$ of t=0.4 and -1.0 is added to $\log_{10} S(k,t)$ of t=3.8.

Fig. 7 (color online) Local width and structure function. (a) Log-log plot of the local width against time. The slope of the dashed line is 0.25. (b) Log-log plot of local structure function against wave number. The solid and dashed lines in (b) have slopes of -3.0 and -2.0, respectively. The time is 3.0. Parameters for the simulation: $\gamma = 3.0$, $\lambda = 2.0$, $\bar{\upsilon}(\bar{r}) = 1.0$.

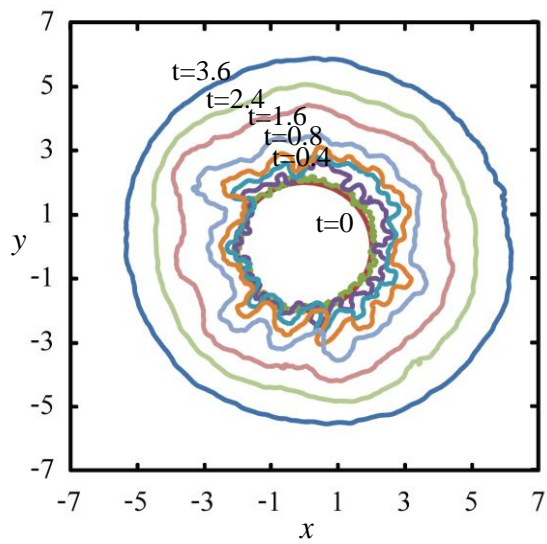

Figure 1 Yong-Jun Chen, Yuko Nagamine, Tomohiko Yamaguchi and Kenichi Yoshikawa

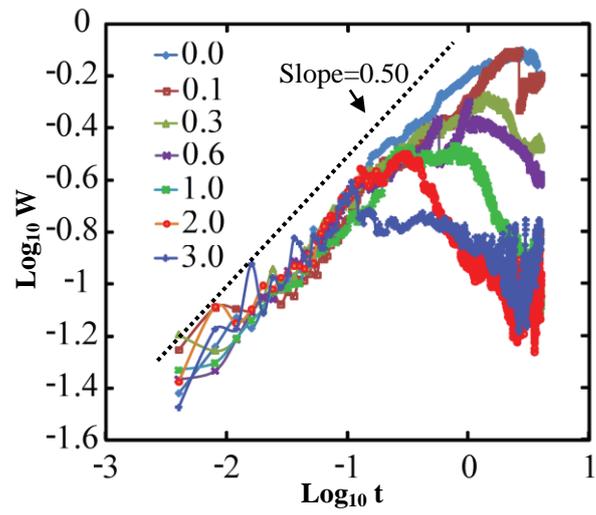

Figure 2 Yong-Jun Chen, Yuko Nagamine, Tomohiko Yamaguchi and Kenichi Yoshikawa

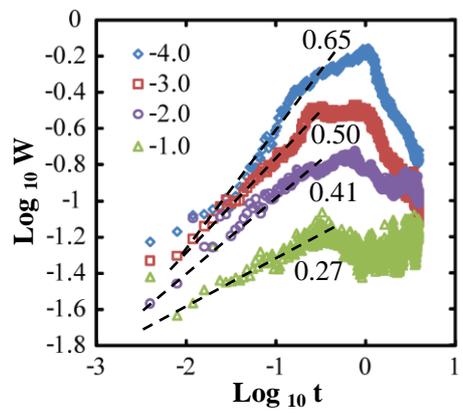

Figure 3 Yong-Jun Chen, Yuko Nagamine, Tomohiko Yamaguchi and Kenichi Yoshikawa

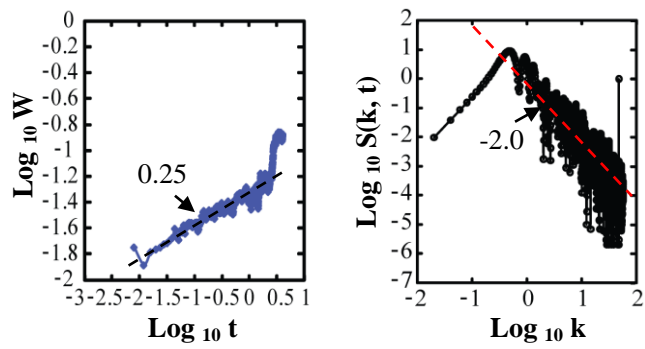

Figure 4 Yong-Jun Chen, Yuko Nagamine, Tomohiko Yamaguchi and Kenichi Yoshikawa

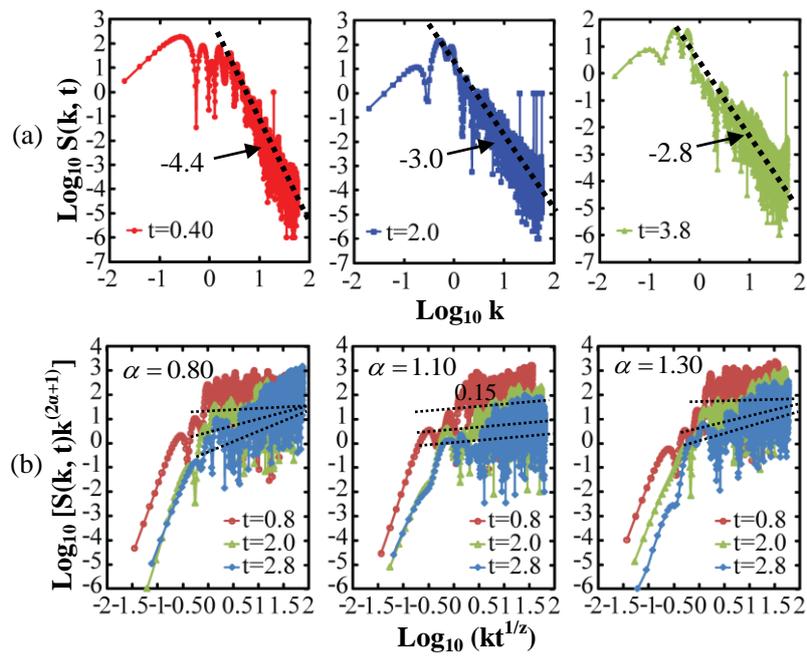

Figure 5 Yong-Jun Chen, Yuko Nagamine, Tomohiko Yamaguchi and Kenichi Yoshikawa

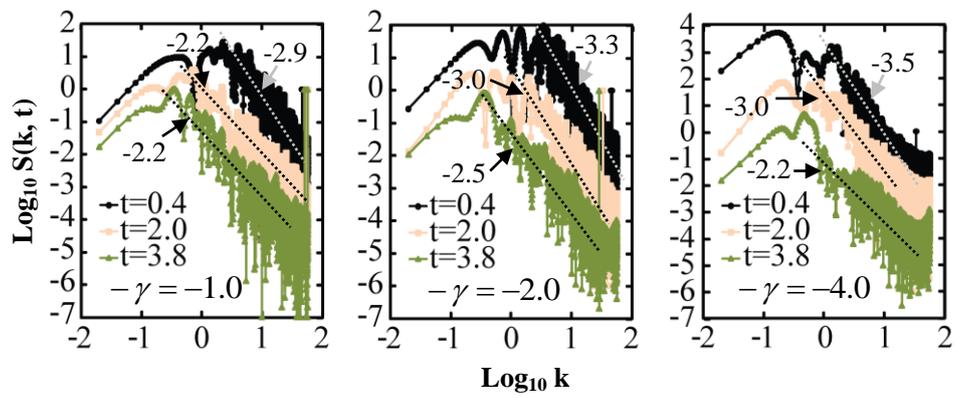

Figure 6 Yong-Jun Chen, Yuko Nagamine, Tomohiko Yamaguchi and Kenichi Yoshikawa

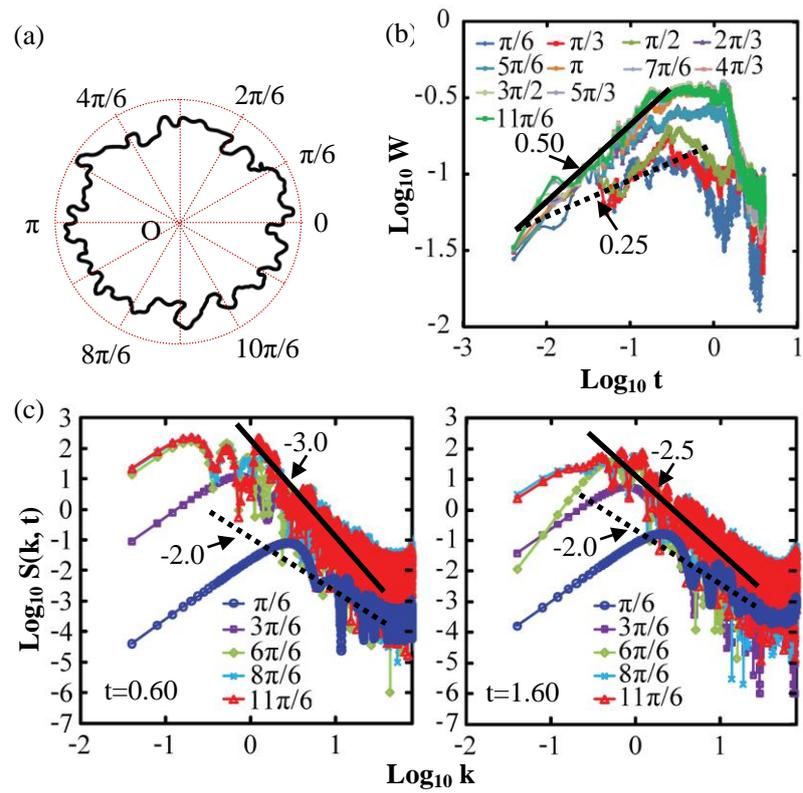

Figure 7 Yong-Jun Chen, Yuko Nagamine, Tomohiko Yamaguchi and Kenichi Yoshikawa